%% file: IEEE-conference-template-062824.tex
\documentclass[conference]{IEEEtran}
\IEEEoverridecommandlockouts %

\pdfoutput=1
\usepackage{multirow}
\usepackage{listings}
\usepackage{xcolor}
\usepackage{caption}
\usepackage{placeins}
\usepackage{url}        %
\usepackage{hyperref}   %
\usepackage{orcidlink}
\usepackage{booktabs}

\usepackage{cite}

\begin{document}

\title{\textsc{ATTest}: \underline{A}gent-Driven \underline{T}ensor \underline{Test}ing for Deep Learning Library Modules}

\author{

\IEEEauthorblockN{
Zhengyu~Zhan\IEEEauthorrefmark{1},
Ye~Shang\IEEEauthorrefmark{1},
Jiawei~Liu\IEEEauthorrefmark{1},
Chunrong~Fang\IEEEauthorrefmark{1},
Quanjun~Zhang\IEEEauthorrefmark{2},
and Zhenyu~Chen\IEEEauthorrefmark{1}
}

\IEEEauthorblockA{
\IEEEauthorrefmark{1}State Key Laboratory for Novel Software Technology,
Nanjing University, Nanjing, China\\
\{522025320208, yeshang\}@smail.nju.edu.cn,
\{jwliu, fangchunrong, zychen\}@nju.edu.cn
}

\IEEEauthorblockA{
\IEEEauthorrefmark{2}School of Computer Science and Engineering,
Nanjing University of Science and Technology, Nanjing, China\\
quanjunzhang@njust.edu.cn
}

}

\maketitle

\begin{abstract}
The unit testing of Deep Learning (DL) libraries is challenging due to complex numerical semantics and implicit tensor constraints. Traditional Search-Based Software Testing (SBST) often suffers from semantic blindness, failing to satisfy the constraints of high-dimensional tensors, whereas Large Language Models (LLMs) struggle with cross-file context and unstable code modifications. This paper proposes \textsc{ATTest}, an agent-driven tensor testing framework for module-level unit test generation. \textsc{ATTest} orchestrates a seven-stage pipeline, which encompasses constraint extraction and an iterative ``generation-validation-repair'' loop, to maintain testing stability and mitigate context-window saturation. An evaluation on PyTorch and TensorFlow demonstrates that \textsc{ATTest} significantly outperforms state-of-the-art baselines such as PynguinML, achieving an average branch coverage of 55.60\% and 54.77\%, respectively. The results illustrate how agent-driven workflows bridge the semantic gap in numerical libraries while ensuring auditable test synthesis. Source code: \url{https://github.com/iSEngLab/ATTest.git}.
\end{abstract}

\begin{IEEEkeywords}
unit test generation, deep learning libraries, LLM agents
\end{IEEEkeywords}

\input{introduction}
\input{related_works}
\input{approach}

\input{demonstration}

\input{evaluation}

\input{conclusion}

\bibliographystyle{IEEEtran}
\bibliography{reference}

\end{document}

%% file: introduction.tex
\section{Introduction}
\label{intro}

Unit testing is essential for software reliability as it provides the feedback loop required for agile development. The automation of this process has been advanced by Search-Based Software Testing (SBST) engines using many-objective algorithms (e.g., EvoSuite~\cite{fraser2011evosuite}, DynaMOSA~\cite{panichella2017automated}, Pynguin~\cite{lukasczyk2022pynguin}) and, more recently, by Large Language Model (LLM)-based frameworks such as LIBRO~\cite{kang2023large} and ChatUniTest~\cite{chen2024chatunitest}. These approaches have substantially reduced the manual effort required to create unit tests for general applications. However, as the software landscape pivots toward data-centric paradigms, specifically within Deep Learning (DL) library like PyTorch and TensorFlow, the nature of the subject under test is fundamentally transformed. This shift reveals three critical limitations in existing methodologies:

First, traditional SBST tools suffer from semantic blindness toward high-dimensional input spaces. Unaware of implicit ``tensor constraints'' (e.g., \texttt{rank}, \texttt{shape}, and \texttt{dtype}), they frequently generate non-compliant inputs that fail at early argument validation, never reaching the core computational logic. Second, existing LLM-based frameworks often struggle with cross-file context, leading to hallucinations or destructive code rewriting where fixing one case inadvertently breaks another. Third, current LLM pipelines (e.g., ChatUniTest~\cite{chen2024chatunitest}) remain rigid; even hybrid techniques like CodaMosa~\cite{lemieux2023codamosa} or neural oracle generators such as ToGA~\cite{dinella2022toga} often cannot adapt to the specialized numerical precision and device-specific state management (e.g., CUDA~\cite{nickolls2008scalable}) required for DL operator testing.

To address these limitations, we introduce \textsc{ATTest}, an Agent-driven Tensor Testing framework designed for DL operators. Instead of treating the LLM as a passive code generator, \textsc{ATTest} models it as an autonomous engineering agent that orchestrates the unit testing lifecycle through a structured workflow.

First, to overcome the semantic blindness of traditional SBST tools in high-dimensional input spaces, \textsc{ATTest} integrates constraint-aware test generation with execution-grounded feedback. Rather than exploring the input space blindly, the agent reasons over implicit tensor constraints and iteratively refines test inputs based on concrete runtime signals, enabling deeper exploration of core numerical logic. Second, to mitigate hallucinations and destructive code rewriting in existing LLM-based frameworks, \textsc{ATTest} introduces a block-level incremental repair mechanism. By organizing generated tests into semantically isolated blocks and restricting modifications to failure-localized regions, the framework preserves consistency and prevents unintended regressions across iterations. Third, in contrast to rigid, hard-coded generation--validation--repair pipelines, \textsc{ATTest} adopts a staged and extensible orchestration workflow. This design allows the agent to dynamically adjust its testing strategy to the specialized rigor of deep learning operator testing, including strict numerical precision requirements and device-specific execution contexts (e.g., CUDA state management~\cite{nickolls2008scalable}), which are insufficiently supported by fixed LLM pipelines.

In summary, this work makes the following contributions:
\begin{itemize}
    \item We propose \textsc{ATTest}, an Agent-driven Tensor Testing framework that integrates constraint-aware generation, execution-driven refinement, and block-level incremental editing to enable stable and auditable unit test synthesis for deep learning operators.
    \item We implement a practical CLI-based toolchain with persistent state management and resumable execution, and evaluate its effectiveness on industrial-grade libraries such as PyTorch and TensorFlow.
\end{itemize}

%% file: related_works.tex
\section{Related Works}
\label{related_works}

Traditional Search-Based Software Testing (SBST) tools (e.g., EvoSuite~\cite{fraser2011evosuite}, Pynguin~\cite{lukasczyk2022pynguin}) maximize coverage but often lack semantic awareness for complex inputs like high-dimensional tensors. To address this, recent approaches extract input constraints from API documentation~\cite{xie2022docter} or mining patterns~\cite{narayanan2023automatic}, integrating them into mutation operators to improve coverage for DL libraries~\cite{krodinger2025constraint}. Concurrently, comprehensive studies demonstrate that LLMs significantly enhance test generation through deep semantic understanding and fine-tuning~\cite{zhang2025large, shang2025large}. To ensure execution correctness, frameworks such as ChatUniTest~\cite{chen2024chatunitest} and TestART~\cite{gu2024testart} establish iterative ``generation-validation-repair'' cycles, while TestSpark~\cite{sapozhnikov2024testspark} incorporates IDE feedback. Furthermore, as evaluation paradigms increasingly emphasize complex, class-level dependencies~\cite{zhang2024testbench}, recent research has shifted toward context-aware, agentic testing. For instance, RATester~\cite{yin2025enhancing} mitigates hallucinations via symbol lookup, and multi-agent systems like QualityFlow~\cite{hu2025qualityflow} and TDFlow~\cite{han2025tdflow} employ specialized roles for planning and debugging to handle intricate, library-specific constraints.

%% file: approach.tex
\section{Approach}
\label{approach}

\begin{figure}[htbp]
    \centering
    \includegraphics[width=\linewidth]{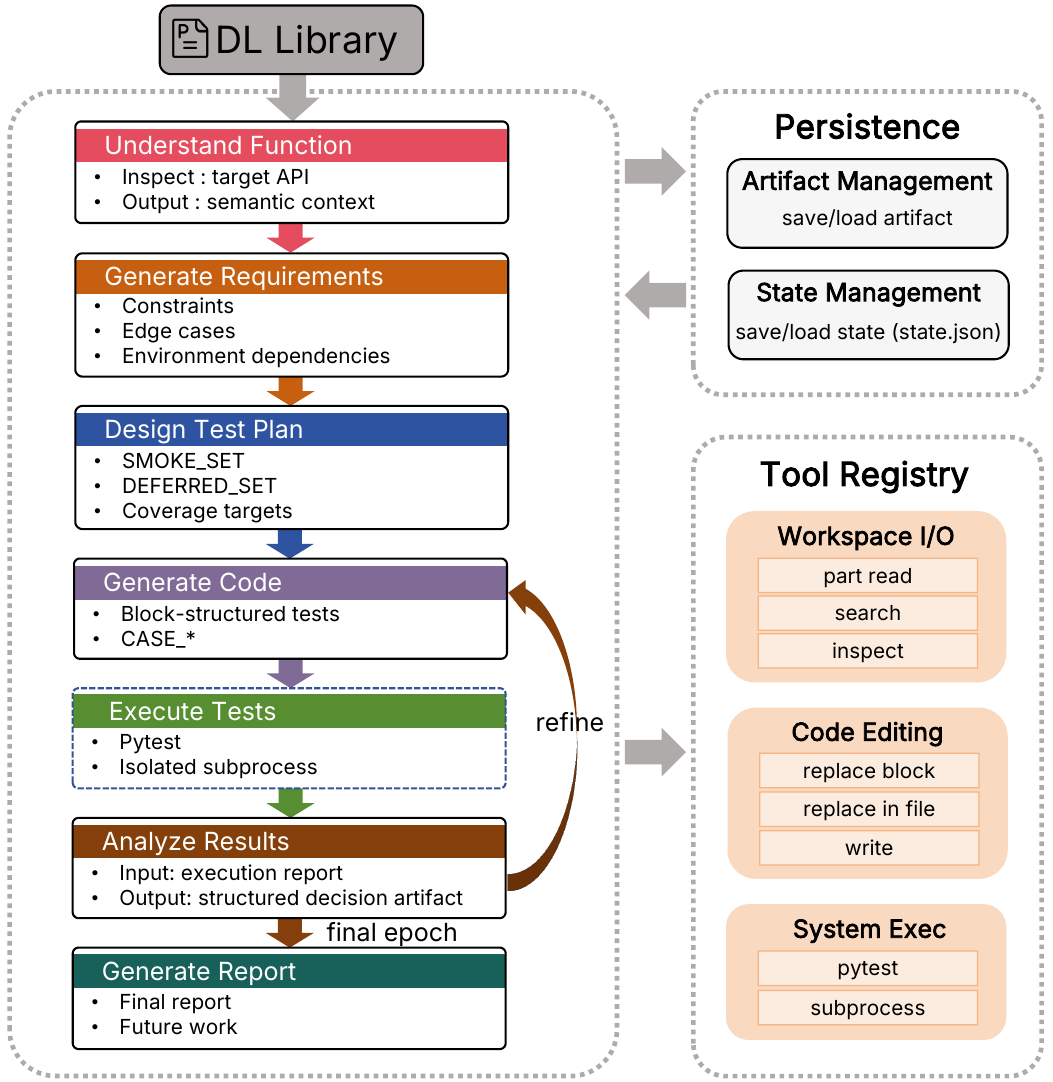}
    \caption{Workflow of \textsc{ATTest}.}
    \label{fig:Workflow}
\end{figure}

\textsc{ATTest} is designed as an autonomous agentic framework for testing complex numerical and DL libraries. Unlike traditional SBST tools that suffer from semantic blindness, \textsc{ATTest} treats the LLM as an autonomous engineering agent that orchestrates the entire unit testing lifecycle through a structured and persistent workflow. The core of our approach is a seven-stage workflow (Figure~\ref{fig:Workflow}) that enables process-level decision-making, iterative refinement, and reproducible execution.

\subsection{Seven-Stage Testing Workflow}
\textsc{ATTest} decomposes module-level unit test generation into a unified seven-stage workflow with clearly defined functional responsibilities. Each stage focuses on a specific engineering objective and processes only task-relevant information, which mitigates context-window saturation and the lost-in-the-middle effect.

Specifically, the workflow consists of the following stages: (1) \emph{Understand Function}, which inspects API signatures, documentation, and source code; (2) \emph{Generate Requirements}, which derives semantic and structural constraints; (3) \emph{Design Test Plan}, which encodes specifications in a machine-readable file (\texttt{test\_plan.json}); (4) \emph{Generate Code}, which synthesizes candidate test cases; (5) \emph{Execute Tests}, which runs generated tests under controlled environments; (6) \emph{Analyze Results}, which interprets execution outcomes and failure signals; and (7) \emph{Generate Report}, which consolidates test artifacts and diagnostics.

This staged decomposition enables \textsc{ATTest} to organize long-horizon testing tasks into manageable units. By isolating semantic reasoning, code synthesis, execution feedback, and reporting into dedicated stages, the framework maintains clarity of responsibilities and supports systematic iteration.

\subsection{Supervisory Orchestration}
To coordinate the execution of the seven-stage workflow, \textsc{ATTest} introduces a dedicated supervisory agent responsible for global planning and process-level control. Rather than following a fixed linear pipeline, this agent dynamically determines stage transitions based on system states, execution feedback, and historical decisions.

At each iteration, the supervisory agent monitors structured artifacts generated by individual stages, including test plans, execution reports, and analysis summaries. Based on these artifacts, it decides whether the workflow should proceed, repeat the current stage, or backtrack to earlier stages for refinement. For example, when failures indicate deficiencies in planning rather than local coding errors, the agent may redirect the workflow from Result Analysis back to Test Plan Design to revise testing objectives.

This orchestration mechanism enables \textsc{ATTest} to support non-linear and adaptive testing processes. By explicitly modeling control flow at the agent level, the framework separates local task execution from global decision-making. This separation allows specialized agents to focus on domain-specific subtasks, while the supervisory agent maintains overall coherence and long-term optimization.

Workflow coordination is further supported by a persistent state management mechanism. All intermediate artifacts, decisions, and tool outputs are serialized into a unified state file. This design enables interrupted workflows to be resumed, failed stages to be retried, and historical decisions to be inspected, thereby improving reproducibility and debuggability.

\subsection{Engineering Mechanisms for Stability}
Long-horizon test generation for numerical libraries is susceptible to instability caused by cascading errors, destructive rewriting, and excessive contextual complexity. To ensure reliable convergence, \textsc{ATTest} incorporates several dedicated engineering mechanisms.

First, \textsc{ATTest} adopts incremental block-level editing. Generated test files are partitioned into semantically isolated blocks (e.g., \texttt{HEADER}, \texttt{CASE\_*}, \texttt{FOOTER}). During each iteration, modifications are restricted to a bounded number of blocks. This policy prevents large-scale rewrites and preserves previously validated test cases.

Second, \textsc{ATTest} employs plan-driven scope management. Test cases are organized into a \texttt{SMOKE\_SET} for rapid validation and a \texttt{DEFERRED\_SET} for deferred exploration. This strategy prioritizes time-to-first-run and controls contextual complexity in iterations.

Third, \textsc{ATTest} applies selective log ingestion during result analysis. Instead of processing complete execution logs, the system extracts only error-relevant fragments using lightweight search and partial-reading tools. This mechanism reduces token consumption and improves repair precision by focusing agent reasoning on diagnostically meaningful information.

%% file: demonstration.tex
\section{Demonstration Scenario}
\label{demo}

We demonstrate the iterative repair capability of \textsc{ATTest} on a representative and constraint-sensitive API, \texttt{torch.nn.utils.spectral\_norm} in PyTorch. This API involves internal tensor reshaping, permutation, and state mutation through buffer and hook registration, making it a typical example of numerical library testing.

\textbf{\textit{Workflow Initialization and Guided Generation.}} The demonstration starts with an interactive workspace initialized through the \textsc{ATTest} CLI. After configuring execution templates and API credentials, the supervisory agent orchestrates the testing workflow under a unified runtime configuration, ensuring consistent collection of failures, tracebacks, and coverage reports.

\begin{figure}[htbp]
\centering
\begin{lstlisting}
{
  "status": "partially_passed",
  "passed": 6,
  "failed": 1,
  "errors": 0,
  "collection_errors": false,
  "block_limit": 3,
  "failures": [
    {
      "test": "TestSpectralNorm.test_invalid_dim_index_exception",
      "block_id": "CASE_12",
      "error_type": "RuntimeError",
      "action": "rewrite_block",
      "note": "A dimension mismatch occurs when dim = -1, where the tensor dimensionality does not match the required permutation ordering."
    }
  ],
  "deferred": [],
  "stop_recommended": false,
  "stop_reason": ""
}
\end{lstlisting}
\caption{Example of an \texttt{analysis\_plan.json}.}
\label{fig:analysis-plan}
\end{figure}

\begin{figure*}[htbp]
  \centering
  \includegraphics[width=0.8\linewidth]{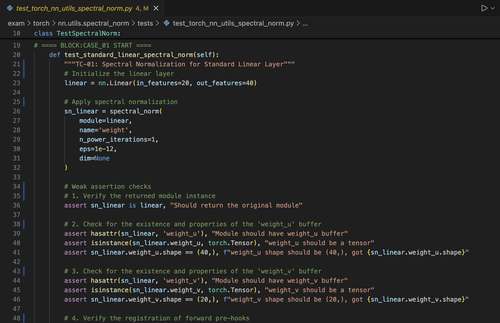}
  \caption{Example of a generated unit test.}
  \label{fig:testcase}
\end{figure*}

In interactive mode, the agent performs function understanding, requirement extraction, and test plan design to establish semantic context. It then generates initial test cases and organizes them into structured code blocks. At designated checkpoints, developers may refine requirements or approve stage transitions while preserving the persistent workflow state.

\textbf{\textit{Multi-Epoch Iterative Repair.}} When initial tests encounter typical tensor dimension and shape constraints, \textsc{ATTest} enters an iterative repair process coordinated by the supervisory agent. In the fourth iteration, the generated test suite contains one failing case, \texttt{TestSpectralNorm.\allowbreak test\_invalid\_dim\_index\_exception}, caused by a mismatch between tensor dimensionality and permutation ordering. Under the same execution configuration, the iteration yields six passing tests, one failure, and 89\% branch coverage.

During result analysis, \textsc{ATTest} selectively extracts error-relevant traceback fragments and synthesizes a structured \texttt{analysis\_plan.json}. This artifact localizes failures to specific code blocks (e.g., \texttt{CASE\_12}) and specifies bounded repair actions (e.g., \texttt{rewrite\_block}) under predefined block limits.
Figure~\ref{fig:analysis-plan} presents an example of the generated \texttt{analysis\_plan.json}, which serves as a compact decision summary and guides subsequent block-level editing.

In the fifth iteration, the repaired test suite converges, with all seven tests passing and overall coverage increasing to 99\%. All intermediate artifacts and execution states are preserved in the workspace for auditing and reproducibility.

\textbf{\textit{Result Consolidation.}} After reaching the user-specified iteration budget, \textsc{ATTest} proceeds to the final report generation stage. The system summarizes achieved coverage, resolved failures, and remaining limitations, and produces a consolidated testing report for developer inspection. Figure~\ref{fig:testcase} illustrates an example of a unit test generated after convergence.

%% file: evaluation.tex
\section{Evaluation}
\label{evaluation}

\textbf{\textit{Experimental Setup.}} We evaluate \textsc{ATTest} on 165 Python modules from two widely used ML libraries, PyTorch and TensorFlow. These modules involve complex API constraints and high cyclomatic complexity, posing substantial challenges to conventional SBST techniques. We adopt PynguinML as a state-of-the-art (SOTA) baseline, as it specifically targets compliant input generation for DL APIs. \textsc{ATTest} is instantiated with the \texttt{deepseek-chat} model. To ensure fairness and reproducibility, all experiments use the same library versions and execution environments as the baseline. Each configuration runs five rounds of iterative test generation and repair.

\textbf{\textit{Structural Coverage.}} We use branch coverage as the primary metric to assess the ability of different tools to explore control-flow paths in the subject under test. Table~\ref{tab:avg-coverage} reports the average branch coverage achieved by \textsc{ATTest} and PynguinML on both libraries.
\textsc{ATTest} consistently outperforms the baseline, achieving average branch coverages of 55.60\% on PyTorch and 54.77\% on TensorFlow, compared to 43.13\% and 39.72\% obtained by PynguinML, respectively. These results indicate that \textsc{ATTest} is more effective in generating structurally valid test inputs under complex tensor constraints.

\begin{table}[htbp]
\centering
\small
\caption{Avg. Branch Coverage Comparison Across Libraries}
\label{tab:avg-coverage}
\resizebox{\linewidth}{!}{
\begin{tabular}{lccc}
\toprule
\textbf{Tool} & \textbf{Library} & \textbf{Avg. Branch Coverage (\%)} & \textbf{Overall Avg. (\%)} \\
\midrule
\multirow{2}{*}{\textsc{ATTest}} & PyTorch & 55.60 & \multirow{2}{*}{55.19} \\
 & TensorFlow & 54.77 & \\
\midrule
\multirow{2}{*}{PynguinML} & PyTorch & 43.13 & \multirow{2}{*}{41.43} \\
 & TensorFlow & 39.72 & \\
\bottomrule
\end{tabular}
}
\end{table}

To normalize performance across modules with varying complexity, we further adopt the relative coverage metric ($cov_{r}$)~\cite{krodinger2025constraint}:
\begin{equation}
cov_{r}(s, e) = \frac{cov(s, e) - \min(cov(s))}{\max(cov(s)) - \min(cov(s))}.
\end{equation}
Here, $s$ denotes a subject module, $e$ denotes one execution configuration, and $cov(s,e)$ represents the achieved branch coverage. Across the evaluated modules, \textsc{ATTest} achieves $cov_r=1.0000$ on a substantial majority of subjects, including core components such as \path{torch.nn.modules.conv} and \path{tensorflow.python.ops.math_ops}. For more challenging modules, such as \path{torch.ao.quantization.quantize} ($cov_r=0.8973$) and \path{tensorflow.python.ops.functional_ops} ($cov_r=0.9781$), \textsc{ATTest} still substantially outperforms the baseline by navigating execution paths that require strict compliance with tensor shape and data type constraints.

\textbf{\textit{Cross-Library Robustness.}} \textsc{ATTest} demonstrates consistent effectiveness across heterogeneous DL libraries. On the TensorFlow benchmark, the tool achieves full relative coverage on 68 modules, reflecting its ability to handle both graph-based and eager execution contexts. Similarly, on PyTorch, \textsc{ATTest} reliably generates compliant \texttt{torch.Tensor} inputs for initialization and functional routines.
The high concentration of $cov_r=1.0000$ scores across both libraries suggests that \textsc{ATTest} generalizes well beyond library-specific implementation details, providing a broadly applicable solution for testing modern numerical software.

%% file: conclusion.tex
\section{Conclusion and Future Work}
\label{conclusion}

This paper presents \textsc{ATTest}, an agent-driven framework that autonomously orchestrates module-level unit test generation for DL libraries. By reframing the testing lifecycle as a structured seven-stage pipeline rather than a passive generation task, \textsc{ATTest} integrates process-level decision-making with persistent state management to overcome intricate tensor constraints and context saturation. Experimental results confirm that this staged orchestration significantly outperforms state-of-the-art baselines in both structural coverage and reliability. Future work will focus on fusing semantic reasoning with symbolic execution, extending the autonomous workflow to library-level integration testing, and adapting the framework to heterogeneous hardware backends.